\documentclass[twocolumn,showpacs,aps,amsmath,amssymb,superscriptaddress,prb]{revtex4-1}
\usepackage{amsfonts}
\usepackage{amsmath}
\usepackage{bm}
\usepackage{graphicx}
\usepackage{booktabs}
\usepackage{overpic}
\usepackage{color}
\usepackage{multirow}

\newcommand{\be}{\begin{equation}}
\newcommand{\ee}{\end{equation}}
\newcommand{\bea}{\begin{eqnarray}}
\newcommand{\eea}{\end{eqnarray}}
\newcommand{\bsube}{\begin{subequations}}
\newcommand{\esube}{\end{subequations}}
\newcommand{\Fig}[1]{Fig.\,\ref{#1}}

\newcommand{\comments}[1]{}

\begin{document}

\title{Molecular Kondo Ball Based on an Endohedral Fullerene}

\author{Xiangyang~Li} \email{xyangli@ustc.edu.cn}
\affiliation{Hefei National Laboratory, University of Science and Technology of China, Hefei, Anhui 230088, China}

\author{Xiao~Zheng} 
\affiliation{Hefei National Laboratory, University of Science and Technology of China, Hefei, Anhui 230088, China}
\affiliation{Department of Chemistry, Fudan University, Shanghai 200438, China}

\author{Bing Wang}
\affiliation{Hefei National Laboratory, University of Science and Technology of China, Hefei, Anhui 230088, China}
\affiliation{Hefei National Laboratory for Physical Sciences at the Microscale, University of Science and Technology of China, Hefei, Anhui 230026, China}

\author{Jinlong~Yang}\email{jlyang@ustc.edu.cn}
\affiliation{Hefei National Laboratory, University of Science and Technology of China, Hefei, Anhui 230088, China}
\affiliation{State Key Laboratory of Precision and Intelligent Chemistry, University of Science and Technology of China, Hefei, Anhui 230026, China}


\begin{abstract}
We theoretically predict a molecular Kondo ball in TiSc$_2$N@C$_{80}$/Al(111), where a Ti-centered $S = 1/2$ local moment is encapsulated within a closed fullerene $\pi$ cage that serves as an itinerant-like electronic reservoir for Kondo screening.
First-principles calculations show that intermediate molecule--substrate hybridization broadens the discrete C$_{80}$ molecular levels into itinerant-like electronic states while preserving the Ti local moment, with similar electronic conditions found for Cd(0001) and In(101).
Hierarchical equations of motion (HEOM) simulations reveal a temperature-dependent Fano resonance with a Kondo temperature of about 60 K.
The key electronic characteristics underlying the Kondo state remain robust against adsorption orientation, demonstrating that the closed cage geometry provides an orientation-tolerant molecular architecture for Kondo screening.

\[
\includegraphics[width=0.55\textwidth]{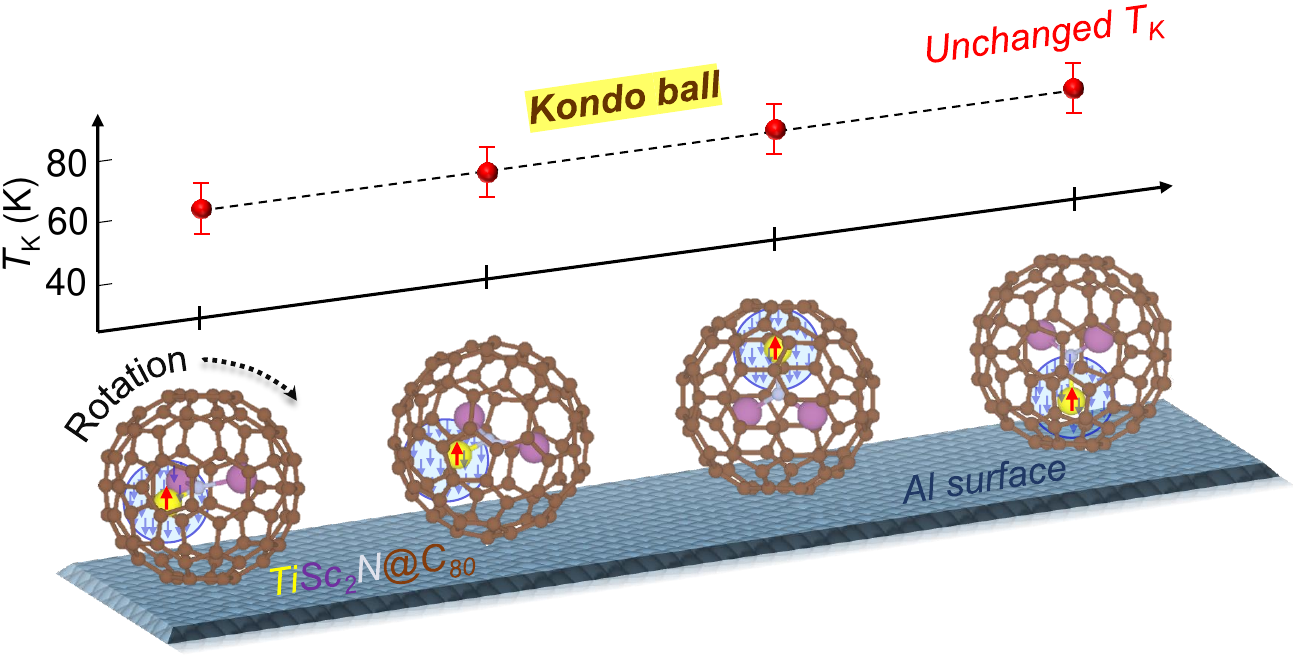}
\]

\end{abstract}

\maketitle

The Kondo effect arises from the many-body screening of a localized magnetic moment by itinerant electrons at low temperatures.\cite{Kon6437,Mad98567}
Surface-adsorbed magnetic molecules provide a chemically tunable platform for studying this effect, because their spin states, orbital symmetries, and molecule--substrate hybridization can be engineered at the molecular level.\cite{Zha051542,Bog08179,Par101370,Sco103560,Li26109,Li202566} 
In most molecular Kondo systems, however, the itinerant electron reservoir is supplied primarily by the metallic substrate, while the molecule mainly hosts the localized spin.\cite{Zha051542,Bog08179,Par101370,Sco103560,Li26109} 
This substrate-dominated screening picture limits the structural programmability of Kondo screening at the molecular scale.

Recent studies have shown that conjugated $\pi$-electron molecules can actively mediate the coupling between localized spins and metallic electronic states, allowing them to function not merely as localized-spin hosts but also as active components of an itinerant-like electronic reservoir.\cite{Odo001549,Boo05267202,Li25086201}
In Co-CoPc complexes adsorbed on Au(111), we introduced a molecular Kondo box, where the planar $\pi$ states of CoPc acquire itinerant-like electronic character through hybridization with Au(111) and screen a Co-localized spin via symmetry-matched coupling to the Co $d_{\pi}$ orbitals.\cite{Li25086201} 
A qualitatively different architecture can arise when the localized spin is placed inside, rather than adjacent to, a closed molecular reservoir. 
In such a geometry, the screening environment surrounds the magnetic center and may therefore become less sensitive to the molecule--surface orientation than in an open molecular framework.

Endohedral fullerenes provide an ideal platform for testing this idea.\cite{Hu242863,Liu18457,Pop104857}
Their carbon cages provide a closed three-dimensional $\pi$-electron framework that spatially surrounds encapsulated metal atoms or clusters, naturally separating a cage-protected localized spin center from an external metallic substrate.
In the isolated-molecule limit, however, the fullerene cage possesses discrete molecular levels rather than a continuous metallic spectrum, and therefore cannot by itself provide an effective itinerant-electron reservoir for spin screening. 
This raises a key question: can interfacial hybridization with a metal surface activate the closed fullerene $\pi$ cage as an itinerant-like electronic reservoir while preserving the encapsulated local moment, thereby realizing a Kondo architecture in which the magnetic impurity is enclosed by its molecular screening environment?
\begin{figure*}[ht]
	\centering
	\includegraphics[width=1.4\columnwidth]{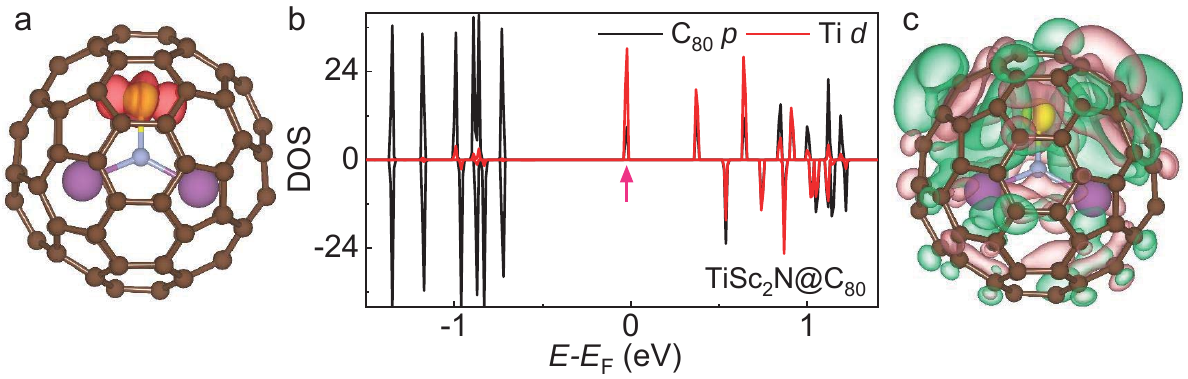}
	\caption{Electronic structure of an isolated TiSc$_2$N@C$_{80}$ molecule. (a) Optimized structure and spin-density distribution of TiSc$_2$N@C$_{80}$.
		The yellow, gray, purple, and brown spheres represent Ti, N, Sc, and C atoms, respectively.
		The red isosurface corresponds to a spin density of 0.04~$e$~\AA$^{-3}$.
		(b) Projected DOS of $\mathrm{C}_{80}$ $p$ orbitals and Ti $d$ orbitals, with the pink arrow marking the SOMO. 
		(c) Real-space wavefunction isosurface of the singly occupied state at an isovalue of $\pm 0.00018$, where cyan and pink denote the positive and negative phases, respectively.}
	\label{fig1}
\end{figure*}

To address this question, we theoretically predict a molecular Kondo ball based on the experimentally synthesized endohedral fullerene TiSc$_2$N@C$_{80}$ adsorbed on Al(111), in which the Ti-centered local moment is enclosed by a substrate-activated fullerene $\pi$-electron reservoir.
First-principles calculations show that an intermediate molecule--substrate hybridization preserves the Ti-centered $S=1/2$ local moment while broadening the originally discrete C$_{80}$ molecular levels into itinerant-like cage states near the Fermi level.
Comparative substrate calculations further reveal that this balanced hybridization is essential: overly strong coupling quenches the local spin, whereas overly weak coupling fails to activate the fullerene cage as an electronic reservoir.
In addition to Al(111), Cd(0001) and In(101) also lie within this intermediate-coupling regime.
Using parameters extracted from the DFT calculations, hierarchical equations of motion simulations\cite{Ye16608,Zha24e1727,Hou15104112,Wan14084713,CuiLei2019,Tan89101} reveal a temperature-dependent Kondo resonance with a characteristic temperature of approximately 60 K.
The relevant electronic structures are insensitive to the adsorption orientation of the C$_{80}$ cage on Al(111), indicating that the electronic ingredients for Kondo screening are retained across the examined configurations.
This work identifies substrate-activated fullerene cages as orientation-tolerant molecular screening environments.
%

\begin{figure*}
	\centering
	\includegraphics[width=1.4\columnwidth]{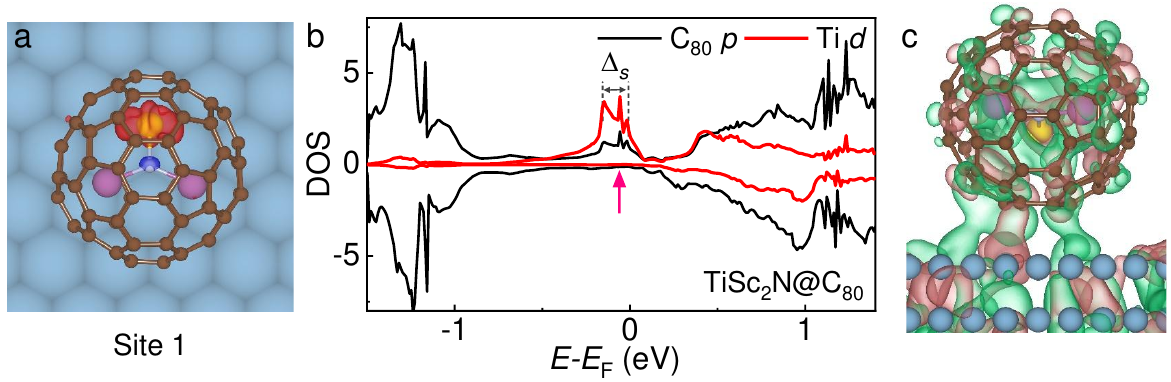}
	\caption{Electronic structure of TiSc$_2$N@C$_{80}$ on the Al(111) surface. 
		(a) Optimized structure and spin-density distribution of TiSc$_2$N@C$_{80}$/Al(111). 
		The red isosurface corresponds to a spin density of 0.04~$e$~\AA$^{-3}$. 
		(b) Projected DOS of the corresponding $\mathrm{C}_{80}$ $p$ orbitals and Ti $d$ orbitals, 
		with the pink arrow marking the SOMO. $\Delta_s$ denotes the splitting width of the singly occupied-state peak.
		(c) Real-space wavefunction isosurface of the SOMO at an isovalue of $\pm 0.00018$, where cyan and pink denote the positive and negative phases, respectively.
	}
	\label{fig2}
\end{figure*}

Figure~\ref{fig1}(a) shows the molecular structure of TiSc$_2$N@C$_{80}$, 
where a TiSc$_2$N mixed-metal nitride cluster with a nearly triangular TiSc$_2$ metal framework is encapsulated inside the C$_{80}$ cage.
This endohedral fullerene has been experimentally synthesized and isolated.\cite{Pop104857,Yan096391}
Low-temperature electron spin resonance (ESR) measurements have established that the unpaired electron is primarily localized on the encapsulated Ti atom,\cite{Pop104857} making TiSc$_2$N@C$_{80}$ particularly suitable for exploring the Kondo screening of a well-defined local moment.
The Ti atom can be formally assigned as Ti(III) with a 3$d^1$ configuration,\cite{Pop104857} corresponding to an $S = 1/2$ spin state.
Consistently, the calculated spin density is strongly concentrated around the Ti atom, as shown by the red isosurface in Figure~\ref{fig1}(a).

For the isolated TiSc$_2$N@C$_{80}$ molecule, the electronic structure consists of a series of discrete molecular levels [Figure~\ref{fig1}(b)].
Its magnetic character is primarily associated with the Ti $d$-dominated singly occupied molecular orbital (SOMO) near the Fermi level, with only weak spin polarization on the C$_{80}$ $p$ orbitals.
Although the SOMO exhibits partial real-space delocalization with $\pi$ character [Figure~\ref{fig1}(c)], it remains an energetically discrete molecular orbital, precluding it from serving as an itinerant-like electronic state.

\begin{table*}[htbp]
	\centering
	\setlength{\tabcolsep}{3.5pt}
	\caption{Substrate-dependent electronic criteria for forming a molecular Kondo ball. ``Yes'' denotes favorable electronic conditions. HEOM calculations are performed only for TiSc$_2$N@C$_{80}$/Al(111) at Site 1.}
	\label{tab:kondo_ball_design_rule}
	\begin{tabular}{cccccc}
		\toprule
		\hline
		\hline
		Type & Ti local spin & C$_{80}$ itinerant-like reservoir & Kondo ball & Representative substrates & $E_{\rm ads}$ (eV) of TiSc$_2$N@C$_{80}$\\
		\hline
		\midrule
		I & Quenched  & Activated & No & Ag(111) & $-$1.85 \\
		II & Quenched  & Activated & No & Au(111) & $-$1.79 \\
		III  & Preserved & Activated & Yes & Al(111) & $-$1.62 \\
		IV  & Preserved & Activated & Yes & Cd(0001) & $-$1.61 \\
		V   & Preserved & Activated & Yes & In(101) & $-$1.34 \\
		VI  & Preserved & Insufficiently activated & No & Sb(111)& $-$0.81  \\
		\hline
		\hline
		\bottomrule
	\end{tabular}
\end{table*}

To activate itinerant-like electronic states associated with the C$_{80}$ cage in TiSc$_2$N@C$_{80}$, while preserving the Ti-centered local spin, we systematically investigate how different metal substrates, including Ag(111), Au(111), Al(111), Cd(0001), In(101), and Sb(111), modulate its electronic structure.
Notably, all these metal substrates are experimentally accessible and can be prepared under realistic experimental conditions.\cite{Zha051542,Li202566,Li235956,Max95R5546,Sha201305,Guo222201311,She22026802}
Theoretical calculations reveal that the interaction strength between TiSc$_2$N@C$_{80}$ and the substrates varies significantly, with the adsorption energy decreasing in magnitude from -1.85 eV on Ag(111) to -0.81 eV on Sb(111); see Table I for details.

For the Ag(111) and Au(111) substrates, strong molecule--substrate coupling can effectively impart itinerant-like electronic character to the C$_{80}$ cage but simultaneously quench the local magnetic moment on the Ti atom (Figs. S1 and S2). 
In contrast, the Al(111), Cd(0001), and In(101) substrates provide an intermediate interaction strength that not only preserves the Ti-centered local magnetic moment but also activates itinerant-like electronic states of the C$_{80}$ cage (Figs. 2, S3 and S4). 
Further comparison shows that although Sb(111) maintains the local spin on the Ti center, its coupling with the molecule is too weak to impart itinerant-like electronic character to the C$_{80}$ cage (Fig. S5). 
These results indicate that the electronic conditions favorable for Kondo screening in TiSc$_2$N@C$_{80}$ depend critically on a delicate balance in molecule--substrate coupling strength.
Overly strong coupling quenches the local magnetic moment, whereas overly weak coupling fails to activate the itinerant-like electron reservoir.
Therefore, an intermediate adsorption energy, for example around -1.5 eV, can simultaneously satisfy the two key requirements of local-moment preservation and itinerant-electron-reservoir activation, thereby defining a substrate-controlled hybridization regime favorable for Kondo ball formation.

Here, TiSc$_2$N@C$_{80}$/Al(111) is selected as a representative system to elucidate the electronic structure underlying Kondo screening. 
\Fig{fig2}(a) shows the optimized adsorption geometry of TiSc$_2$N@C$_{80}$ on Al(111) at Site 1, together with the corresponding spin-density distribution.
The unpaired electron remains mainly localized on the Ti atom, indicating that the Ti-centered $S=1/2$ local moment is preserved, similar to that in the isolated molecule [\Fig{fig1}(a)].

In sharp contrast to the isolated molecule, the projected density of states (DOS) of C$_{80}$ $p$ and Ti $d$ orbitals in the adsorbed configuration exhibits pronounced splitting and broadening [\Fig{fig2}(b)]. 
In particular, the C$_{80}$ $p$ states form a quasi-continuous distribution around the Fermi level, suggesting that the Al(111) substrate activates the fullerene cage into an itinerant-like electronic reservoir for screening the Ti local moment.
Meanwhile, the Ti $d$ orbital below the Fermi level develops into a finite-width spectral resonance with an energy width of $\Delta_s$, reflecting its hybridization with the reservoir states.

The real-space distribution of the SOMO further supports this picture [\Fig{fig2}(c)]. 
Rather than being confined to the Ti center, the SOMO is delocalized over the TiSc$_2$N@C$_{80}$ framework and extends across the molecule--substrate interface, providing evidence of substantial hybridization between TiSc$_2$N@C$_{80}$ and Al(111).
Such interfacial hybridization endows the C$_{80}$ cage with itinerant-like electronic character and provides the microscopic basis for Kondo screening.
This spatial organization differs from the conventional substrate-dominated molecular Kondo geometry.\cite{Mad98567,She22026802,Che2535464,Liu269650,Che262856}

%
\begin{figure}[htbp]
	\centering
	\includegraphics[width=1\columnwidth]{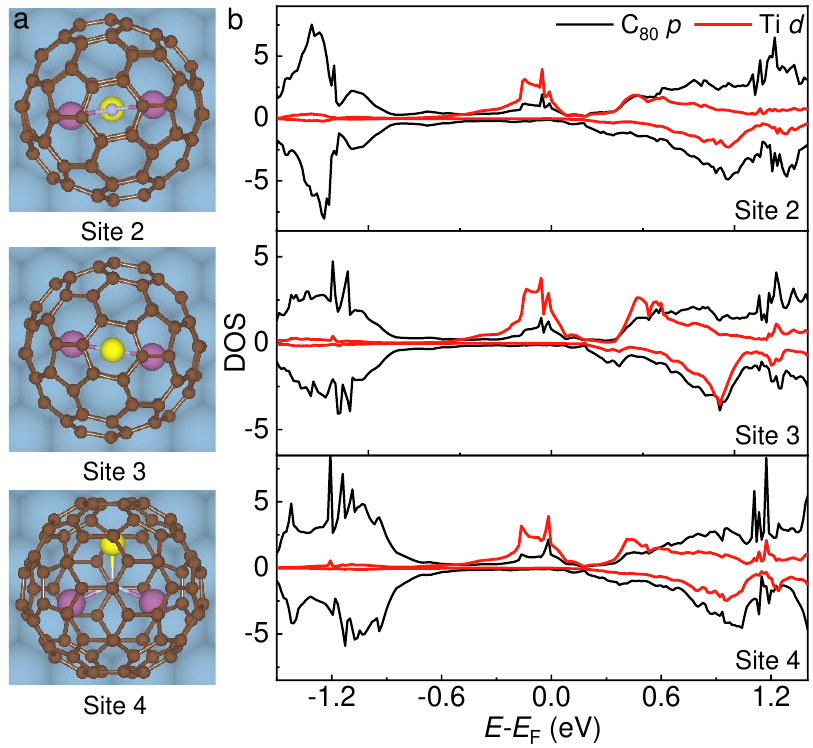}
	\caption{Electronic structures of TiSc$_2$N@C$_{80}$ on the Al(111) surface in different adsorption configurations.
		(a) Optimized structural models.
		(b) Projected density of states of Ti $d$ orbitals and C$_{80}$ $p$ orbitals for the corresponding configurations.}\label{fig3}
\end{figure}

To further examine the robustness of the electronic structure of TiSc$_2$N@C$_{80}$ on Al(111), we consider four distinct adsorption configurations.
As shown in \Fig{fig3}(a), the other three configurations (Sites 2--4), together with the Site 1 configuration discussed above, describe the rolling of the C$_{80}$ cage on the Al(111) surface, thereby sampling different adsorption sites and molecular orientations.
The projected DOS of the Ti $d$ orbitals and C$_{80}$ $p$ orbitals is then compared for these configurations.
As shown in \Fig{fig3}(b), the overall line shapes of both the Ti $d$ and C$_{80}$ $p$ states are well preserved, indicating that the Ti-centered local moment and the itinerant-like reservoir character of the C$_{80}$ cage are retained across different adsorption geometries.
Remarkably, the splitting width $\Delta_s$ of the Ti $d$ spectral peak below the Fermi level remains nearly unchanged among the four configurations.
These results demonstrate that the key electronic ingredients for fullerene-cage-mediated spin screening are robust against variations in adsorption site and molecular orientation, consistent with the closed-cage screening geometry and substantially relaxing the need for precise control over molecular adsorption.

Based on the above analysis, one representative adsorption configuration (Site 1) is selected to construct a single-orbital Anderson impurity model (SIAM) to describe the tip/TiSc$_2$N@C$_{80}$/Al(111) junction.\cite{And6141,Keh961} 
Within the model, the Ti-centered local moment is treated as the magnetic impurity and described by
\be\label{AIM}
H_{\rm imp} = \epsilon_d(\hat{n}_{\uparrow} + \hat{n}_{\downarrow}) + U\hat{n}_{\uparrow}\hat{n}_{\downarrow},
\ee
where $\epsilon_d$ denotes the energy of the impurity level and $U$ is the on-site Coulomb repulsion.
The itinerant-like electronic states associated with the C$_{80}$ cage ($s$) and the scanning tunneling microscopy (STM) tip ($t$) are treated as two electron reservoirs.
Their couplings to the Ti impurity are characterized by the hybridization functions
$\Gamma_{\alpha}(\omega) = \frac{\Delta_{\alpha}}{2}\frac{W_\alpha^2}{(\omega-\Omega_\alpha)^2 + W_\alpha^2}$, where $\alpha=s,t$.
The relevant model parameters, including $\epsilon_d$, $U$, $\Delta_{\alpha}$, $\Omega_{\alpha}$, and $W_{\alpha}$, are extracted from the DFT results.
In particular, the determination of the hybridization parameter $\Delta_s$ is illustrated in \Fig{fig2}(b).

\begin{figure}
	\centering
	\includegraphics[width=1\columnwidth]{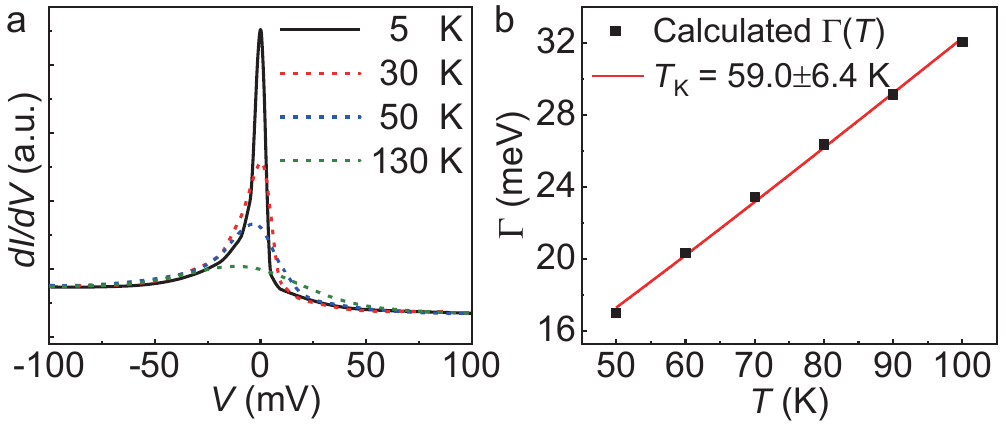}
	\caption{Kondo effect in TiSc$_2$N@C$_{80}$/Al(111). 
		(a) Temperature-dependent $dI/dV$ spectra obtained from HEOM simulations. 
		(b) Evolution of the half width at half maximum (HWHM) of the Kondo resonance peak as a function of temperature.
	}\label{fig4}
\end{figure}

The $dI/dV$ spectra of TiSc$_2$N@C$_{80}$/Al(111) are calculated using the HEOM method.\cite{Ye16608,Hou15104112,Wan14084713,CuiLei2019} As shown in \Fig{fig4}(a), a pronounced peak emerges near the Fermi level at $T=5$ K. With increasing temperature, this peak is gradually suppressed and broadened, which is characteristic of a Kondo resonance.
To quantify the spectral evolution, the simulated $dI/dV$ spectra are fitted using the Fano line shape,\cite{Fan611866}
\begin{equation}
	\frac{dI}{dV}
	\simeq
	a\frac{(\varepsilon+q)^2}{1+\varepsilon^2}+b,
	\label{eq}
\end{equation}
where $\varepsilon=\frac{V-V_0}{\Gamma}$.
Here, $V_0$ denotes the center of the differential conductance peak, $\Gamma$ is the half width at half maximum (HWHM), $q$ is the Fano asymmetry parameter, and $a$ and $b$ are fitting parameters.\cite{Fan611866,Mir102257}
The temperature-dependent HWHM values extracted from the Fano fits are shown in \Fig{fig4}(b).

To extract the characteristic energy scale of the Kondo resonance, $\Gamma$ is analyzed as a function of temperature. 
For an $S=1/2$ Kondo system, the temperature dependence of the resonance width in the low-temperature regime can generally be described by Fermi-liquid theory:
\begin{equation}
\Gamma(T)=\sqrt{(\pi k_{\rm B}T)^2+2(k_{\rm B}T_{\rm K})^2},
\end{equation}
where $k_{\rm B}$ is the Boltzmann constant and $T_{\rm K}$ is the Kondo temperature.
The red solid line in \Fig{fig4}(b) shows the fitting of $\Gamma(T)$ using the Fermi-liquid relation.
Within the examined temperature range, the extracted Kondo temperature of TiSc$_2$N@C$_{80}$/Al(111) is $T_{\rm K}=59.0\pm6.4$ K. 
This energy scale is lower than that previously reported for Co@CoPc/Au(111) ($T_{\rm K}=123$ K),\cite{Li25086201} indicating relatively weaker yet clearly resolved Kondo screening in the present system.

From an experimental perspective, the proposed TiSc$_2$N@C$_{80}$/Al(111) platform appears feasible.
TiSc$_2$N@C$_{80}$ has already been experimentally synthesized and isolated.\cite{Pop104857,Yan096391}
Related nitride clusterfullerenes, including Sc$_3$N@C$_{80}$ and Er$_3$N@C$_{80}$, have been deposited onto metal surfaces under ultrahigh-vacuum conditions and characterized by STM and scanning tunneling spectroscopy (STS).\cite{Nor08052080}
Moreover, STM experiments on C$_{60}$/Al(111) demonstrate the experimental accessibility of fullerene adsorption on a clean Al(111) surface.\cite{Joh98314}
These established procedures suggest that purified TiSc$_2$N@C$_{80}$ could be thermally deposited at low coverage onto clean Al(111) to realize the proposed Kondo ball geometry.

The predicted Kondo state could be probed by low-temperature STM/STS through the temperature dependence of the near-zero-bias Fano resonance, from which the Kondo temperature $T_{\rm K}$ can be extracted.\cite{Mad98567}
The observation of Kondo signatures arising from localized magnetic moments associated with interstitial V atoms in VTe$_2$ provides experimental precedent showing that spatially embedded spins can remain detectable.\cite{Liu198572}
The main challenge in the present system is that the Ti spin is buried inside the C$_{80}$ cage, such that it must be probed indirectly through the cage states.
Temperature-dependent spectroscopy combined with spatially resolved $dI/dV$ mapping could help distinguish the cage-mediated Kondo resonance from ordinary molecular features.

\section{Conclusions}

In summary, we have established TiSc$_2$N@C$_{80}$/Al(111) as a molecular Kondo ball, in which a Ti-centered $S=1/2$ local moment is encapsulated within a substrate-activated fullerene screening environment.
Comparative first-principles calculations identify an intermediate molecule--substrate hybridization regime that preserves the Ti local moment while rendering the C$_{80}$ cage itinerant-like, whereas stronger or weaker coupling disrupts this balance.
The key electronic characteristics underlying the Kondo state are robust against adsorption geometry, while HEOM simulations for the representative configuration yield $T_{\rm K}=59.0\pm6.4$ K.
More broadly, the Kondo ball architecture places a cage-protected local moment inside a finite molecular screening reservoir, establishing a distinct route to molecular Kondo physics in which the geometry of the reservoir itself becomes a design degree of freedom.

\section{acknowledgments}
This work is supported by the National Natural Science Foundation of China (22573109) and the Innovation Program for Quantum Science and Technology (2021ZD0303306).
The computational resources were provided
by the Supercomputing Center of the University of Science and Technology of China.

\section{Competing interests}
The authors declare no competing interests.

\section{Additional information}
Supplementary information is available for this paper at https://doi.org.

\bibliographystyle{aip}
\bibliography{bibrefs}

\end{document}